\documentclass[]{emulateapj}
\shorttitle{On MHD turbulence}
\advance\textwidth by 1mm
\advance\columnsep by -1mm
\begin{document}
\title{Scaling laws and diffuse locality of balanced and imbalanced MHD turbulence.}
\author{A. Beresnyak, A. Lazarian}
\affil{Dept. of Astronomy, Univ. of Wisconsin, Madison, WI 53706}
\email{andrey, lazarian@astro.wisc.edu}

\def\L{{\Lambda}}
\def\l{{\lambda}}

\begin{abstract} 
The search of ways to generalize the theory of strong MHD
turbulence for the case of non-zero cross-helicity (or energy
imbalance) has attracted considerable interest recently. In
our earlier publications we performed three-dimensional numerical
simulations and showed that some of existing models are
inconsistent with numerics. In this paper we focused our attention
on low-imbalance limit and performed new high-resolution
simulations. The results strongly suggest that in the limit of small
imbalances we smoothly transition to a standard Goldreich-Sridhar (1995)
balanced model. We also claim that Perez-Boldyrev (2009) model that
predicts the same nonlinear timescale for both components
due to so-called ``dynamic alignment'' strongly contradicts
numerical evidence.
\end{abstract}

\keywords{MHD -- turbulence -- ISM: kinematics and dynamics}

\section{Introduction}
MHD turbulence has attracted attention of astronomers since mid 1960s.
As most astrophysical media are ionized, plasmas are coupled to the magnetic fields
\citep[see, e.g.,][]{biskamp}. A simple one-fluid description known
as magnetohydrodynamics or MHD
is broadly applicable to most astrophysical environments on macroscopic scales.
On the other hand, turbulence has been observed in various circumstances 
and with a huge range of scales \citep[see, e.g.,][]{armstrong,chepurnov}.

As with hydrodynamics which has a ``standard'' phenomenological model of
energy cascade (Kolmogorov, 1941), MHD turbulence has one too. This is
the Goldreich-Sridhar
model \citep[henceforth GS95]{GS95} that uses a concept of critical balance,
which maintains that turbulence will stay marginally strong down the cascade.
The spectrum of GS95 is supposed to follow a $-5/3$ Kolmogorov scaling.
However, a shallower slopes has been reported by numerics, which motivated
to modify GS95 (see, e.g., Boldyrev 2005, 2006, Gogoberidze 2007).

The other problem of GS95 is that it is incomplete, as it does not treat
the most general imbalanced, or cross-helical case. As turbulence is a stochastic
phenomenon, an average zero cross helicity does not preclude a fluctuations
of this quantity in the turbulent volume. Also, most of astrophysical turbulence
is naturally imbalanced, due to the fact that it is generated by a strong
localized source of perturbations, such as the Sun in case of solar wind
or central engine in case of AGN jets.

Several models of imbalanced turbulence appeared recently:
\citet{lithwick2007} henceforth LGS07, \citet{BL08},
\citet{chandran2008}, \citet{PB09} henceforce PB09, \citet{podesta}.
The full self-consistent
analytical model for strong turbulence, however, does not yet exist. In this situation
observations and direct numerical simulations (DNS) of MHD turbulence will provide
necessary feedback to theorists.
We concentrated on two issues,
namely that a) the energy power-law slopes of MHD turbulence can not be measured directly
from available numerical simulations, supporting an earlier claim in BL09b,
b) the ratio of energy dissipation rates is a very robust quantity that can be used
to differentiate among many imbalanced models. We believe, that taken together
they resolve confusion related to the subject.


In what follows in
\S2 we briefly describe numerical methods,
\S3 we show and discuss dissipation rates as the most robust measures
in numerical turbulence, in \S4 we argue that it is impossible to measure
asymptotic spectral slopes of turbulence directly from currently available DNS,
in \S5 we discuss the perspectives of using DNS to constrain models,
in \S6 we present our conclusions.

\section{Numerical setup}

We solved incompressible MHD or Navier-Stokes
equations:

\begin{equation}
\partial_t{\bf w^\pm}+\hat S ({\bf w^\mp}\cdot\nabla){\bf w^\pm}=-\nu_n(-\nabla^2)^{n/2}{\bf w^\pm}+ {\bf f^\pm},
\end{equation}

where $\hat S$ is a solenoidal projection and ${\bf w^\pm}$ (Elsasser variables)
are ${\bf w^+=v+b}$ and ${\bf w^-=v-b}$ where we use velocity $\bf{v}$
and magnetic field in velocity units ${\bf b=B}/(4\pi \rho)^{1/2}$.
Navier-Stokes equation is a special case of equations (1), where $b\equiv 0$
and, therefore, both equations are equivalent when $w^+\equiv w^-$.
The RHS of this equation includes a linear dissipation term which
is called viscosity or diffusivity for $n=2$ and hyper-viscosity
or hyper-diffusivity for $n>2$ and the driving force ${\bf f^\pm}$.
The special case of ${\bf f^+}={\bf f^-}$ is a velocity driving.
We solved these equations with a pseudospectral code that was described
in great detail in our earlier publications BL09a, BL09b. Table~1 enumerates
latest high-resolution runs, which
were performed in so-called reduced MHD approximation, where the ${\bf w^\pm}$ component
parallel to the mean field (pseudo-Alfv\'en mode) is omitted and so are
the parallel gradients in the nonlinear term ($(\delta {\bf w^\mp}\cdot\nabla_\|)\delta {\bf w^\pm}$. Under these assumptions one studies purely Alfv\'enic dynamics
in a strong mean field, i.e., Alfv\'enic turbulence. 

\begin{table}
\caption{}
  \begin{tabular}{c c c c c c}
    \hline\hline
Run  & Resolution & $f$ & Dissipation & $\epsilon^+/\epsilon^-$ & $(w^+)^2/(w^-)^2$   \\
   \hline
B1 &  $1024\cdot 3072^2$ & $w^\pm$ & $-3.3\cdot10^{-17}k^6$ & $\sim 1$ & $\sim 1$ \\

B2 &  $768\cdot 2048^2$ & $v$ & $-3.1\cdot10^{-16}k^6$ &  $\sim 1$ & $\sim 1$  \\

B3 &  $768\cdot 2048^2$ & $w^\pm$ & $-3.1\cdot10^{-16}k^6$ &  $\sim 1$ & $\sim 1$  \\

B4 &  $768\cdot 2048^2$ & $w^\pm$ & $-6.7\cdot10^{-5}k^6$ &  $\sim 1$ & $\sim 1$ \\

I1 &  $512\cdot 1024^2$ & $w^\pm$ & $-1.9\cdot10^{-4}k^2$ & 1.187 &  $1.35\pm 0.04$ \\

I2 &  $768^3$ & $w^\pm$ & $-6.8\cdot10^{-14}k^6$ & 1.187 &  $1.42 \pm 0.04$   \\

I3 &  $512\cdot 1024^2$ &$w^\pm$ & $-1.9\cdot10^{-4}k^2$ & 1.412 &  $1.88\pm 0.04$   \\

I4 &  $768^3$ & $w^\pm$ & $-6.8\cdot10^{-14}k^6$ &  1.412 & $1.98\pm 0.03$ \\

I5 &  $1024\cdot 1536^2$ & $w^\pm$ & $-1.5\cdot10^{-15}k^6$ &   2   &    $5.57\pm 0.08$  \\

I6 &  $1024\cdot 1536^2$ & $w^\pm$ & $-1.5\cdot10^{-15}k^6$ & 4.5   &  $45.2\pm1.5$   \\

   \hline

\end{tabular}
  \label{experiments}
\end{table}

\section{Nonlinear cascading and dissipation rate}
One of the most robust quantity in numerical simulations of MHD turbulence
is the energy cascading rate or dissipation rate. In high-Reynolds number
turbulence energy has to cascade through many steps before dissipating and
the dissipation is negligible on the outer (large) scale. Therefore,
nonlinear energy cascading rate and dissipation rate are used interchangeably.


In hydrodynamic turbulence the dissipation rate and the spectrum of velocity
are connected by the well-known Kolmogorov constant\footnote{This equation
is subject to intermittency
correction, see, e.g. \citet{frisch}, which is not particularly
relevant for our discussion.}:

\begin{equation}
E(k)=C_K \epsilon^{2/3} k^{-5/3}.
\end{equation}

The important fact that strong hydrodynamic turbulence dissipates in one
dynamic timescale $l/v$ is reflected by $C_K$ being close to unity ($\sim 1.6$).
In MHD turbulence, however, there two energy cascades
(or ``Elsasser cascades'') and there two
dissipation rates, $\epsilon^+$ and $\epsilon^-$. The question of how these
rates are related to velocity-like Elsasser amplitudes ${\bf w}^+$ and ${\bf w}^-$
 is one of the central questions of imbalanced MHD turbulence.
Each model of strong imbalanced turbulence
advocates a different physical picture of cascading
and provides a different relation between
the ratio of energies $(w^+)^2/(w^-)^2$ and ratio of fluxes $\epsilon^+/\epsilon^-$.

\begin{figure}
\includegraphics[width=0.98\columnwidth]{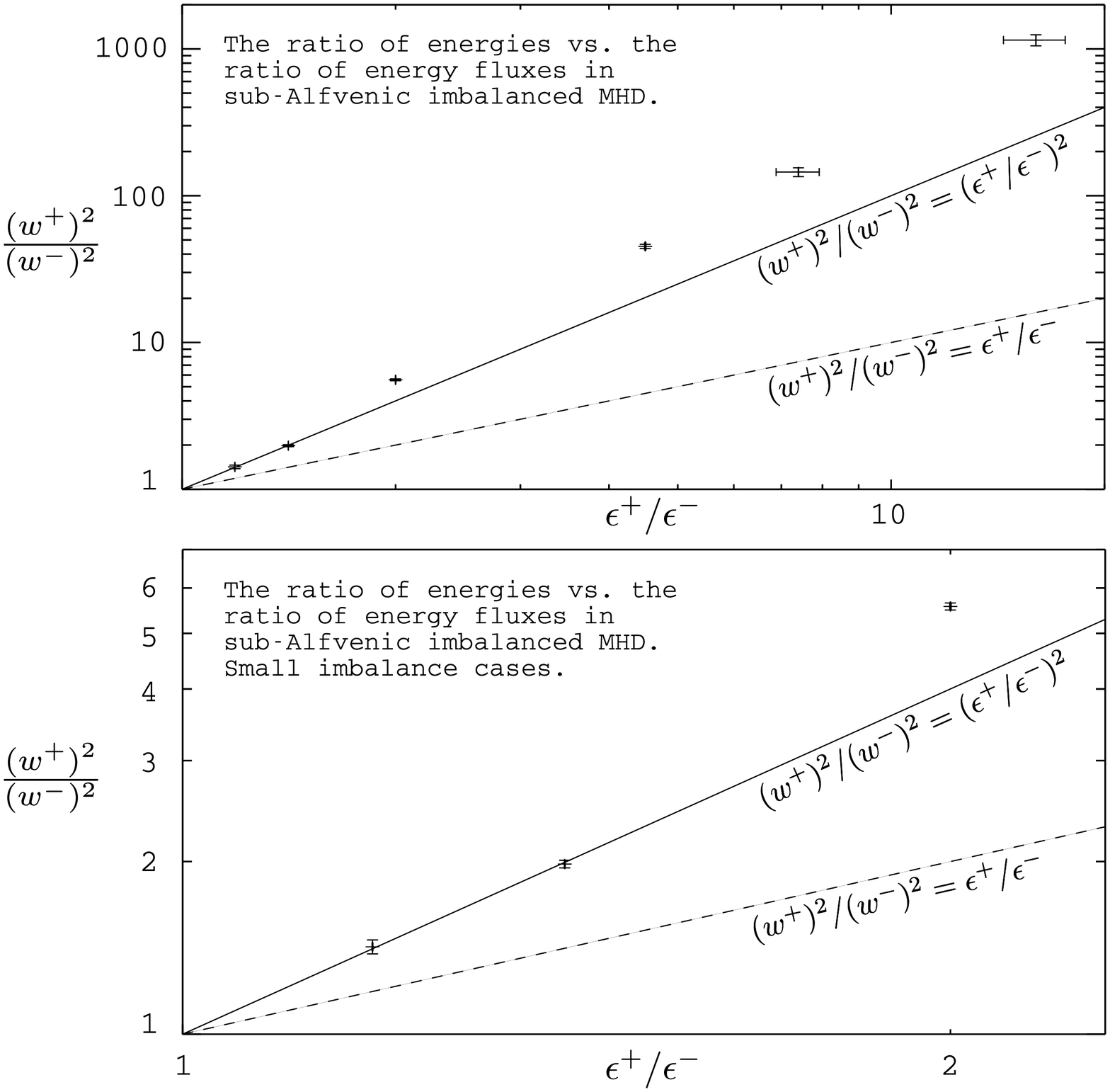}
\caption{Energy imbalances versus dissipation rate imbalance.
Lower panel shows a magnified portion of the
upper panel. Solid line: LGS07 prediction, dashed line: a formula from PB09,
this also is a prediction for purely viscous dissipation.
The point indicate measurements from simulations, where errorbars indicate
fluctuation in time. On this plot I1 and I3 are omitted as they are close
to I2 and I4. I1 and I3 are simulations with normal viscosity which have
slightly lower energy imbalance than I2 and I4, see Table 1.
This is an indication that in these simulations viscosity was affecting outer scales. 
Two high imbalance points are taken from BL09a. 
For a fixed dissipation ratio the energy imbalance has
a tendency to only {\it increase} with resolution.}
\label{dissip}
\end{figure}

Goldreich-Sridhar model (GS95) predicts that in the balanced case
the cascading is strong and each wave is cascaded by the shear rate
of the opposite wave, i.e.,

\begin{equation}
\epsilon^+=\frac{(w^+_l)^2w^-_l}{l},\ \ \ \ \epsilon^-=\frac{(w^-_l)^2w^+_l}{l}.
\end{equation}

It is similar to Kolmogorov cascade with $w$'s replacing $v$. Although
this model does not make predictions for the imbalanced case, one could hope
that in the case of small imbalance these formulae will still work.
In this case we will obtain $(w^+)^2/(w^-)^2=(\epsilon^+/\epsilon^-)^2$.
LGS07 argued that this relation will hold even for large imbalances.

For the purpose of this short paper we mostly discuss the prediction of LGS07,
$(w^+)^2/(w^-)^2=(\epsilon^+/\epsilon^-)^2$ and the prediction of PB09
that nonlinear timescales are equal for both waves, which effectively lead to
\footnote{Both of these predictions are
subject to intermittency corrections. We average $(w^+)^2$ and $(w^-)^2$
over volume and time. This averaging does not take into account possible
fluctuations in $\epsilon^+$ and $\epsilon^-$. We believe, however, that these
effects are small, as long as we use the second-order measures, such as energy.
The issue of $\epsilon^+$ and $\epsilon^-$ fluctuations
is investigated further in \citet{BVL10}.} $(w^+)^2/(w^-)^2=\epsilon^+/\epsilon^-$
(see corresponding equation in PB09, $w^+/w^-=\sqrt{\epsilon^+/\epsilon^-}$).
Note, that the last prediction is also true for highly viscous
flows
($Re=Re_m\ll 1)$. It could be rephrased that PB09 predicts turbulent
viscosity which is equal for both components.

Compared to spectral slopes, dissipation rates are robust quantities that require
much smaller dynamical range and resolution to converge. Fig. \ref{dissip} shows
energy imbalance $(w^+)^2/(w^-)^2$ versus dissipation rate imbalance $\epsilon^+/\epsilon^-$ for simulations I2, I4, I5 and I6. We also use two data points from
our earlier simulations with large imbalances, A7 and A5 from BL09a.
I1 and I3 are simulations with normal viscosity similar to I2 and I4.
They show slightly less energy imbalances than I2 and I4 (Table~1).

We see that most data points are above the line which is the prediction
of LGS07. In other words, one can deduce that numerics strongly suggest
that 

\begin{equation}
\frac{(w^+)^2}{(w^-)^2}\geq \left(\frac{\epsilon^+}{\epsilon^-}\right)^2.
\end{equation} 

Although there is a tentative correspondence between LGS07
and the data for small degrees of imbalance, the
deviations for large imbalances are significant.
Also, the numerics suggests that in the case
of small imbalances the cascading smoothly transition to the
balanced case, i.e. GS95 model, while in the case of strong imbalance
it suggests that the strong component cascading rate is smaller
than what is expected from strong cascading.

As to PB09 prediction, it is inconsistent with data
for all degrees of imbalance including those with small imbalance
and normal viscosity, i.e. I1 and I3.

\begin{figure}
\includegraphics[width=1.0\columnwidth]{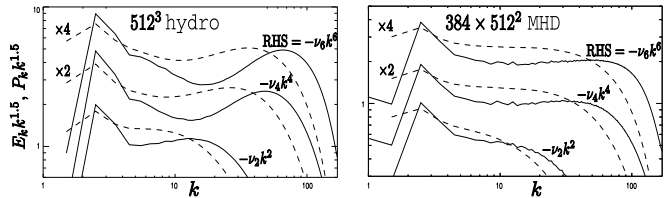}
\caption{The bottleneck effect in MHD and hydro turbulence. Left panel: hydro simulations with different order $n$, of the linear dissipation written as $-\nu_n(-\nabla^2)^{n/2}$, right panel: same for MHD turbulence.
Solid lines show $E_k$, the conventional 3D spectra integrated over solid angle,
while dashed lines show $P_k=2\int_k^\infty E_{k'}dk'/k'$ which is a Fourier transform
of a structure function. For more detail on $E_k$ and $P_k$ see BL09b.}
\label{bottle}
\end{figure}

\section{On the spectral slopes and diffuse locality of MHD turbulence.}

Most of the attention of the theory
has been directed towards the self-similar (or approximately self-similar) regime between
dissipation and driving which is colloquially known as ``inertial range'' (Kolmogorov, 1941).

Although attempts to study this regime started long time ago, it it not until recently
when simulations with resolution higher than $256^3$ has become commonplace. At this
point the interpretation of numerical simulations of MHD turbulence has been
strongly affected by experience obtained from hydrodynamic simulations.

In hydrodynamics, the correspondence with Kolmogorov slope has been fairly
elusive\footnote{See, e.g., \citet{chen97} for $\sim -1.3$ slope for passive scalar in
Kolmogorov turbulence.} for about a decade, even though the same
simulations produced Kolmogorov constants which were close to what has been
observed earlier in experiments. The fact that such correspondence has been
found in simulations with relatively small, less than a thousand,
Reynolds numbers, demonstrated that hydrodynamic turbulence is fairly
local and in order to reproduce asymptotic cascading only a few
steps in log-k space is necessary. The energy slopes, however, were affected
by the bottleneck effect.


In MHD turbulence, the observed flat (power-law) energy spectra has been
prematurely interpreted as ``inertial range''. However, as it turned out,
the flat spectra of MHD turbulence is an indication of a {\it lack}
of a good inertial range, rather than its presence.
Indeed, in simulations with hyperdiffusion we compared hydrodynamic and MHD slopes
and found that while hydrodynamic energy spectra are highly distorted by bottleneck
effect, the MHD spectra stay very flat (see Fig. \ref{bottle}).

As it is not known a-priory what is the contribution of the {\it systematic error}
of the spectral slope measurement that comes from bottleneck effect, it is therefore
impossible to measure true asymptotic slopes directly. Also, it is incorrect to claim
that bottleneck effect is absent in simulations with second-order (natural) viscosity,
as the existence of the effect was clearly demonstrated
in numerical simulations \citep{kaneda}.

Fig.~\ref{bottle} shows a comparison between hydrodynamic and MHD energy slopes
in $512^3$ simulations. As we see, the spectra show a variety of bottleneck
effects, depending on the order of viscosity and type of
simulation (MHD or hydro). Also, there are two types of spectrum, $E_k$ and $P_k$
(see BL09b) and while $E_k$ is used in most numerical papers, it is $P_k$, a
Fourier transform of the structure function which is directly predicted from
Kolmogorov model. While in the asymptotic regime of exact power-law scaling,
$P_k$ and $E_k$ has the same slope, in a realistic numerical simulation
they differ quite a lot. From Fig.~\ref{bottle} it is not immediately obvious
that MHD slopes
are shallower than hydro slopes. Most of the publications that made aforementioned
claim had performed only MHD simulations and compared MHD slope
with asymptotic Kolmogorov slope, i.e. $-5/3$.

\begin{figure}
\plotone{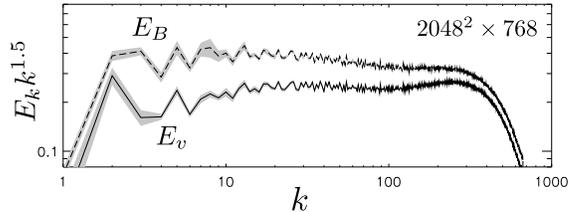}
\caption{Kinetic and magnetic spectra from B3. We show this spectrum
rather that slightly-higher resolution B1 because it had much longer
time evolution. We see that magnetic and kinetic spectra have slightly
different amplitudes and slopes. This is an indication that even with
this resolution one does not have a precise asymptotic regime
of MHD turbulence in a strong field. A weak bottleneck effect is noticeable,
which was absent in lower resolution runs (BL09b). This bottleneck effect
is a hint that we are finally starting to see locality of MHD turbulence.}
\label{2048}
\end{figure}

Our study BL09b reported that MHD turbulence is less
local than hydrodynamic turbulence. This is clearly demonstrated by a) the
lack of visible bottleneck effect in MHD turbulence, while it was clearly present
in hydro turbulence (Fig.~\ref{bottle}), b) the dependence of kinetic and magnetic spectra on driving. Indeed, in case with Elsasser driving magnetic energy dominates by 20-30\% (see Fig. \ref{2048}),
while for velocity driving this is not the case.

\begin{figure}
\begin{center}
\includegraphics[width=0.7\columnwidth]{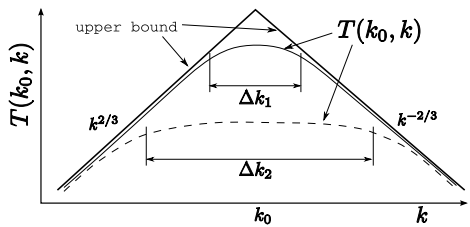}
\end{center}
\caption{A cartoon of energy transfer window $T(k_0,k)$
\citep[for definition, see][]{eyink09} which
has upper bounds $k^{2/3},\,k^{-2/3}$ from theory, due to constraints
on $\delta w^\pm_l$.
However, this upper bound, in practice, could be consistent with both rather
``local'' transfer (upper solid curve) or ``non-local'' or ``diffuse-local''
transfer (lower dashed curve). For more study of energy transfer from simulations
see \citet{BVL10}.}
\label{nonlocality}
\end{figure}

An analytical bound for nonlocality can be obtained through H\"olger inequality
and scalings of the turbulent fields \citep{eyink09}. This bound is shown on Fig.~\ref{nonlocality}. From practical viewpoint, however, this bound does not set a strict
constraint on the ``width'' of the energy transfer window, $T(k_0,k)$, which describe
the energy transfer between wavevectors $k_0$ and $k$, as the maximally
efficient transfer at $k_0$ could still be much lower that the estimate provided
by H\"olger inequality. We conclude that, from practical standpoint,
MHD turbulence can still be ``diffuse local'' i.e. less local
than hydrodynamic turbulence despite this analytical bound.

\section{Discussion}
Although in this short paper we mostly relied on robust quantities, such
as total energies and dissipation rates, we believe that numerical simulations
have a wealth of data to be analyzed by theorists. One of the most important
measures not mentioned in this paper is the anisotropy of MHD turbulence.
It had been considered in great detail in our earlier publication BL09a.
In particular, we refer the reader to the result of BL08, BL09a that the
anisotropy of strong component is smaller than the anisotropy of weak component.
This fact is inconsistent with both the naive application of GS95 critical balance
(which would have predicted the opposite), or the derivation in LGS07 that
suggests that the both waves have the same anisotropy.
We believe it is incorrect to rely only on one particular measure obtained
from simulations, such as the slope of a particular type of spectrum
of the {\it total} energy (see Fig.~\ref{2048} for difference
between magnetic and kinetic spectra, or Fig.~\ref{bottle} for difference between
$E_k$ and $P_k$.) to compare theories and simulations.

PB09 claims that the nonlinear timescales for both components are equal, i.e. 
there is a turbulent viscosity which is the same for both components, regardless
of the degree of imbalance. This seems counter-intuitive for transition
to freely-propagating Alfvenic waves (i.e. infinite imbalance). The formula
in PB09, $w^+/w^-=\sqrt{\epsilon^+/\epsilon^-}$ suggests that the asymptotic
($Re=Re_m \gg 1 $) prediction for energy imbalance in this case will be the same
as in highly viscous case ($Re=Re_m \ll 1$),
i.e. $(w^+)^2/(w^-)^2=\epsilon^+/\epsilon^-$.
This is at odds with numerical evidence,
which suggests $(w^+)^2/(w^-)^2\geq(\epsilon^+/\epsilon^-)^2$.

PB10 claimed that the disagreement between their model and numerics in BL09a
is only exhibited in highly imbalanced simulations.
This is incorrect. In fact, BL09a studied a range of imbalances
$\gamma=w^+/w^-$ starting with 1 (balanced case) and also 2, 10 and a 30.
All of them, including one with small imbalance, showed significant inconsistencies
with PB09 model. Also, as we showed in this paper, the asymptotic regime
of very small imbalances show the same inconsistency with PB09.
Unfortunately, PB10 only dealt with
the issue of the spectral slope,
which is notoriously difficult to measure.
We, however, believe that the total dissipation rates present a more robust measure
and provides an acid test for any local theory of imbalanced turbulence.

In BL09a we discovered and described in detail an empirical fact that
one cannot simulate large imbalances with normal (n=2) dissipation.
This empirical fact also suggests that the strong
component, $w^+$ in our notation, have much larger nonlinear timescale
than the weak component $w^-$. This is supported directly by the time evolution of 
$w^+$ (Fig. 4 of BL09a) and by the shorter inertial range
of $w^+$ (Fig. 10 of BL09a).
In a subsequent publication, PB10 have found a similar empirical evidence,
in particular they claim that a resolution of $1024$ is required to simulate
imbalances of $\gamma \sim w^+/w^-\sim 2$ and one needs to increase the resolution
by a factor of $\gamma^\beta$ where $\beta$ is $2/3$ or $3/4$.
However, this empirical fact directly contradicts to the claim of PB09 model
that both timescales are the precisely the same, due to ``dynamic alignment''.
As PB09 predicts the same nonlinear timescales for both components, they must
have {\it the same} dissipation cutoffs. This is contrary with what is observed
in numerics.

Motivated by longer nonlinear timescales of the strong component we
used hyperdiffusion in this paper, as well as BL09a. The dissipation
for $n=6$ hyperdiffusion is a steep function of wavenumber (cutoff scales
as $\nu^{-1/(s+n+1)}$, where $s$ is the slope), this allowed us to ensure
that dissipation is kept fairly close to Nyquist frequency even for strong
component, thus allowing us to simulate large imbalances. 



\section{Conclusion}
In this paper we mostly studied the regime of small imbalances in MHD turbulence.
In this regime imbalanced MHD turbulence is similar
to the balanced MHD turbulence in a sense that the cascading
will be similar to one in GS95 model (or LGS07), i.e., both waves will be cascaded
strongly and the ratio of energies will be determined by
$(w^+)^2/(w^-)^2=(\epsilon^+/\epsilon^-)^2$.
Larger imbalances show that
$(w^+)^2/(w^-)^2>(\epsilon^+/\epsilon^-)^2$, which suggests that
weak component does not have enough amplitude to provide strong cascading
for opposing strong component (BL08). 
Also we show that PB09 model that claim the same nonlinear
timescales for both components due to ``dynamic alignment''
strongly contradicts numerical evidence.

\acknowledgments
AB thanks IceCube project for support of his research and TeraGrid
for computational resources.
AL acknowledges the  NSF grant AST-0808118 and support from
the Center for Magnetic Self-Organization.

\end{document}